\documentclass[pre,twocolumn,floatfix]{revtex4-1}

\usepackage{natbib}
\usepackage{url}
\usepackage{bm}
\usepackage{graphicx}
\usepackage{amsmath}
\usepackage{amssymb}
\usepackage{amscd}
\usepackage{epstopdf}
\usepackage{verbatim}
\usepackage[usenames]{color}
\usepackage{array}
\usepackage{xspace}
\usepackage{xstring}
\usepackage[normalem]{ulem}
\usepackage{xr}
\usepackage[breaklinks=true,
            pdfborder={0 0 1},
            colorlinks=true,
            linkcolor=black,
            citecolor=blue,
            urlcolor=blue]{hyperref}

\newcommand{\ie}{\textit{i.e.}\xspace}
\newcommand{\eg}{\textit{e.g.}\xspace}

\definecolor{Blue}{rgb}{0,0,0.8}
\definecolor{BlueB}{rgb}{0,0,0.5}
\definecolor{Red}{rgb}{0.8,0,0}
\definecolor{RedB}{rgb}{0.6,0,0}
\definecolor{Green}{rgb}{0,0.5,0}
\definecolor{GreenB}{rgb}{0,0.8,0}
\definecolor{Purple}{rgb}{1,0,1}

\newcommand{\kt}{k_\mathrm{B}T}
\newcommand{\Rnp}{R_\text{np}}
\newcommand{\nb}{n_\text{b}}
\newcommand{\epstar}{{\epsilon^*}}
\newcommand{\thetac}{\theta_\text{d}}

\newcommand{\nf}{N_\text{faces}}
\newcommand{\ffree}{f_\text{free}}
\newcommand{\fbind}{f_\text{b}}
\newcommand{\Nrod}{N_\text{rods}}
\newcommand{\frod}{{\Delta f}}
\newcommand{\fperim}{f_\text{perim}}

\newcommand{\sigmab}{\sigma_\text{b}}
\newcommand{\Usphere}{U_\text{sphere}}
\newcommand{\Ucyl}{U_\text{cyl}}
\newcommand{\Utot}{U_\text{tot}}
\newcommand{\Ntot}{N_\text{tot}}
\newcommand{\nmin}{n_\text{min}}

\begin{document}

\graphicspath{{figures/}}

\title{Faceted particles formed by the frustrated packing of anisotropic colloids on curved surfaces}

\author{Naiyin Yu}
\thanks{These authors contributed equally}
\author{Abhijit Ghosh}
\thanks{These authors contributed equally}
\author{Michael F Hagan}

\email{hagan@brandeis.edu}

\affiliation{Martin Fisher School of Physics, Brandeis University, Waltham, MA, USA.}

\begin{abstract}
We use computer simulations and simple theoretical models to analyze the morphologies that result when rod-like particles end-attach onto a curved surface, creating a finite-thickness monolayer aligned with the surface normal. This geometry leads to two forms of frustration, one associated with the incompatibility of hexagonal order on surfaces with Gaussian curvature, and the second reflecting the deformation of a layer with finite thickness on a surface with non-zero mean curvature. We show that the latter effect leads to a faceting mechanism. Above threshold values of the inter-particle attraction strength and surface mean curvature, the adsorbed layer undergoes a transition from orientational disorder to an ordered state that is demarcated by reproducible patterns of line defects. The number of facets is controlled by the competition between line defect energy and intra-facet strain. Tuning control parameters thus leads to a rich variety of morphologies, including icosahedral particles and irregular polyhedra. In addition to suggesting a new strategy for the synthesis of aspherical particles with tunable symmetries, our results may shed light on recent experiments in which rod-like HIV GAG proteins assemble around nanoscale particles.
\end{abstract}

\maketitle

\section{Introduction}
The formation of crystalline or liquid crystalline order on curved surfaces is frustrated, resulting in topological defects. For example, while the densest packing of disks is a six-fold coordinated hexagonal lattice which perfectly tiles the plane, placing such a lattice on a spherical surface requires at minimum 12 five-fold coordinated disks \cite{Thompson1904,Kleman1989}. Similarly, placing a nematic phase of rodlike molecules on a spherical surface requires defects with net topological charge +2 \cite{Poincare1885,Mermin1979}.
Despite intense theoretical and experimental research into ordering on curved surfaces for more than 100 years (\eg \cite{Thompson1904,Bowick2000,Mermin1979,Bowick2002,Kleman1989,Caspar1962,Nelson2002,Altschuler1997,Bausch2003,Irvine2010,Fernandez-Nieves2007,Lopez-Leon2012,Lopez-Leon2011,Liang2011,Liang2013,Lopez-Leon2011a,Meng2014,Kusumaatmaja2013,Irvine2012,Irvine2012a,Grason2013,Azadi2014,Azadi2016,Mbanga2012,Bruss2012,Grason2013a,Grason2015}), studies have focused on systems in which molecules are confined within the local tangent surface.

Here, we study the coupling between particle shape and curvature that arises when anisotropic (rod-like) particles tend to align along the surface normal.
Our model is motivated by recent experiments on the assembly of rod-like HIV GAG proteins assembling around spherical nanoparticles \cite{Goicochea2011}, and the discovery that rodlike molecules with interparticle attractions can form stable monolayers \cite{Barry2010,Baker2010a,Yang2012}. We find that this geometry leads to two forms of frustration, one familiar from the packing of disks on a sphere, and the second reflecting the strain required to deform a layer of finite thickness on a curved surface.  Frustrated sphere packing is an effect of intrinsic geometry (metric in origin) and thus requires non-zero Gaussian curvature. In contrast, the strain associated with layer thickness arises due to coupling between bend and splay and depends on extrinsic (mean) curvature. Thus, the latter effect occurs even in cylindrical geometries with zero Gaussian curvature. The interplay between frustrations of extrinsic and intrinsic origin leads to a rich variety of faceted shapes, including those with icosahedral symmetry and other polyhedra with lower symmetry, whose features can be readily tuned by controlling template size and inter-particle interactions. In addition to elucidating how multiple forms of packing constraints can conspire to modulate order, our findings identify a new design strategy for synthesizing particles with aspherical shapes.  Such particles are of great interest in the nanomaterials and colloidal communities, since they can assemble into an extraordinary range of structures (e.g. \cite{Anders2014,Glotzer2007a,Singh2007,Duguet2011,Sacanna2010a,Yi2013,Yang2013,Whitelam2014,Wang2012,Wang2014,Walther2013}). Previous studies have shown that flexible ligands on isolated or clustered nanoparticles can order to form spherically asymmetric coatings\cite{Miller2011,Larson-Smith2011, Harkness2011,Ghorai2010,Ghorai2007,Gentilini2010,Gentilini2009,Cesbron2012,Carney2008,Yu2012a,VanLehn2013,Tung2013,Tawfick2012,Stewart2012,Singh2011,Possocco2012,Pons-Siepermann2012,Pons-Siepermann2012a,Singh2007,Sknepnek2012,Jadhao2014,Vernizzi2007,Santos2012,Badia1997,Frederick2011,Landman2004,Lane2010,Luedtke1998,Luedtke1996,Widmer-Cooper2014,Yang2010a,Asai2015,Koch2014}. The current work demonstrates that rigid ligands can also order to form diverse morphologies that, despite their complexity, can be predicted from simple statistical mechanics arguments.

\begin{figure}[h]
\centering
\includegraphics[width=0.4\textwidth]{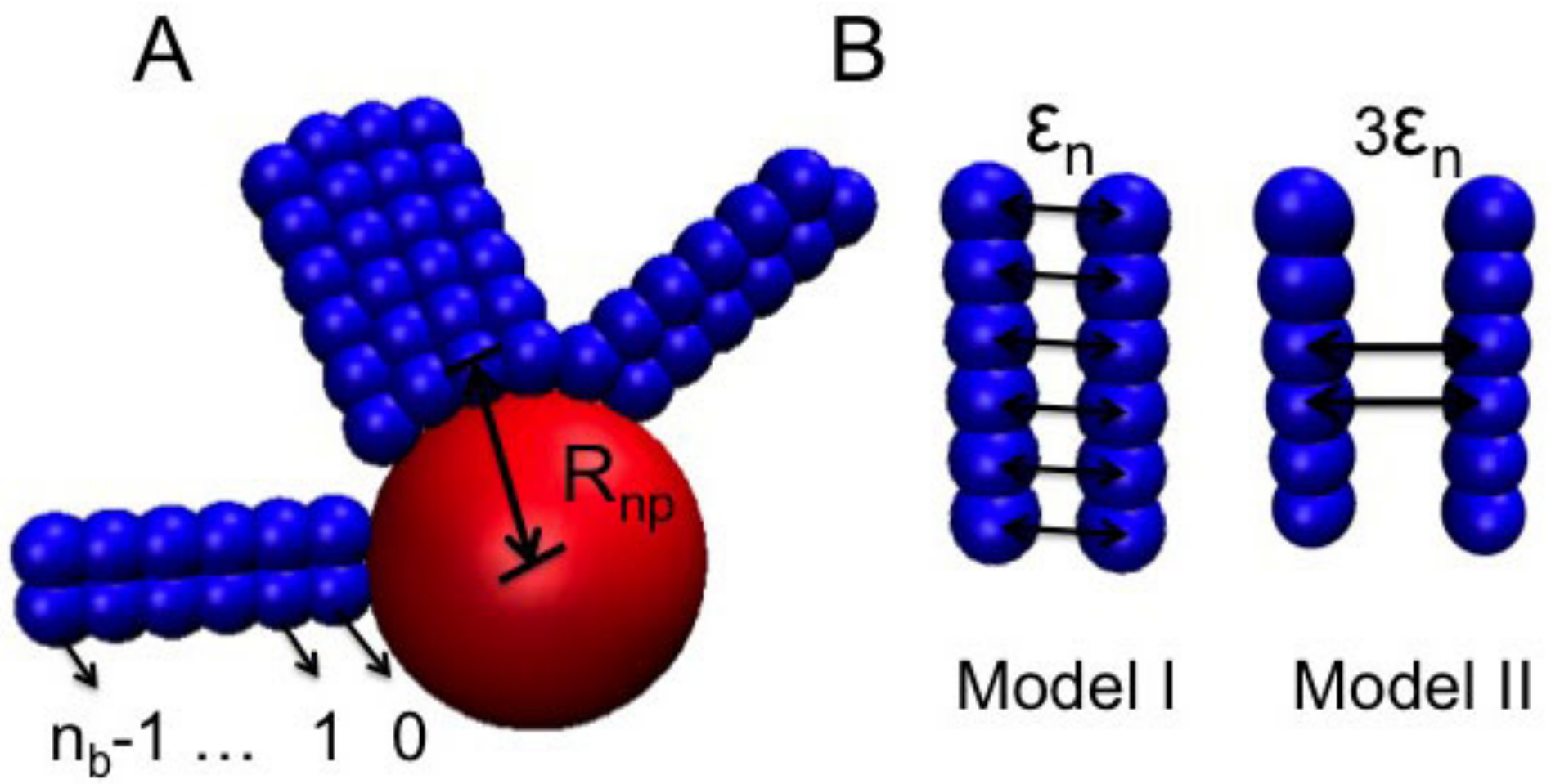}
\caption{Schematic of the models. {\bf (A)} Rods are modeled as rigid strings of beads. Examples are shown with $\nb = 6$ beads. The bottom bead in each rod is attracted to the nanoparticle. The distance between the centers of the nanoparticle and the first bead of an adsorbed rod is denoted by $\Rnp$. {\bf (B)} Two inter-rod  potentials are considered. In Model I, bead-bead attractions are distributed uniformly along the rods; in Model II, interactions occur only between the middle two beads, with the bead-bead interaction strength $\epsilon$ increased to maintain the same net rod-rod interaction strength.}
\label{fig:schematic}
\end{figure}

\section{Model and methods}
\label{sec:model}
\textbf{Spherical nanoparticle in 3D.}
  Our system contains a rigid nanosphere with radius $\Rnp -\sigmab/2$ and $\Nrod$ rigid rods comprised of $\nb$ beads separated by a distance equal to their diameter $\sigmab$. Rods dynamically adsorb end-first onto the nanosphere, and are driven by lateral rod-rod interactions to form a layer aligned with the local nanosphere normal(Fig.~\ref{fig:schematic}).  Beads from different rods experience excluded volume interactions represented by a repulsive Lennard-Jones potential $V_\text{LJ}(r;\sigmab)$, Eq.~\ref{eq:LJ}. In addition, beads with the same index (counting from $n=0$ at the rod bottom) in nearby rods experience attractive interactions represented by a Morse potential $V_\text{morse}(r;\epsilon_n,\sigmab,\alpha=7.5/\sigmab)$, Eq.~\ref{eq:morser}, with $\epsilon_n$ as the well-depth for the attraction between beads with index $n$:
\begin{align}
V_\text{LJ}(r;s)=&\left[\left(\frac{s}{r}\right)^{12} - 1\right]H(s-r)
\label{eq:LJ} \\
V_\text{morse}(r;\epsilon,s,\alpha)=& \epsilon \left(e^{-2 \alpha(r-s) } -2 e^{-\alpha (r-s)} \right) \nonumber \\
& \times H(1.5s-r) -V_\text{shift}
\label{eq:morser}
\end{align}
  where $H(x)$ is the Heaviside function and $V_\text{shift}\approx -0.001$.  We consider two distributions of attractions along the rods (Fig.~\ref{fig:schematic}): (Model I) uniform attractions with  $\epsilon_n=\epsilon\ \forall n$ , and (Model II) attractions occur for the middle 1/3 of beads, with strength $\epsilon_n=3 \epsilon$.
 All beads experience excluded volume interactions with the nanoparticle given by $V_\text{LJ}(r;\Rnp)$ with $\Rnp$ the center-to-center distance between the nanosphere and a bead on its surface. In addition, the first bead in each rod ($n=0$) is attracted to the nanoparticle by $V_\text{morse}(r;\epsilon_\text{np},\Rnp,\alpha=7.5/\sigma_b)$ with $\epsilon_\text{np}=20 \kt$ parameterizing the bead-nanoparticle interaction strength.

\emph{Simulations.} We performed Brownian dynamics simulations with HOOMD \cite{Anderson2008, nguyen2011, LeBard2012}. The simulations used a set of non-dimensional units, described in Ref.~\cite{HOOMDman:2009}.  In this article, we present all energies and lengths in units of the thermal energy $\kt$ and the bead diameter $\sigmab$ respectively.

Simulation initial conditions were representative of a rapid quench from high temperature. Such a configuration was obtained in two steps. First, rod configurations and orientations were chosen randomly except that excluded volume overlaps between pairs of rods or a rod and the nanoparticle were not allowed. Then, dynamics were integrated for $\sim 5\times10^6$ time steps with attractive interactions disabled ($\epsilon=\epsilon_\text{np}=0$), but repulsive interactions operational.  Once this initial condition was generated, attractive interactions were turned on, and the simulation was integrated for at least $2\times10^9$ time steps, with time step size $10^{-3}$ dimensionless time units.

\begin{figure}
\centering
\includegraphics[width=0.3\textwidth]{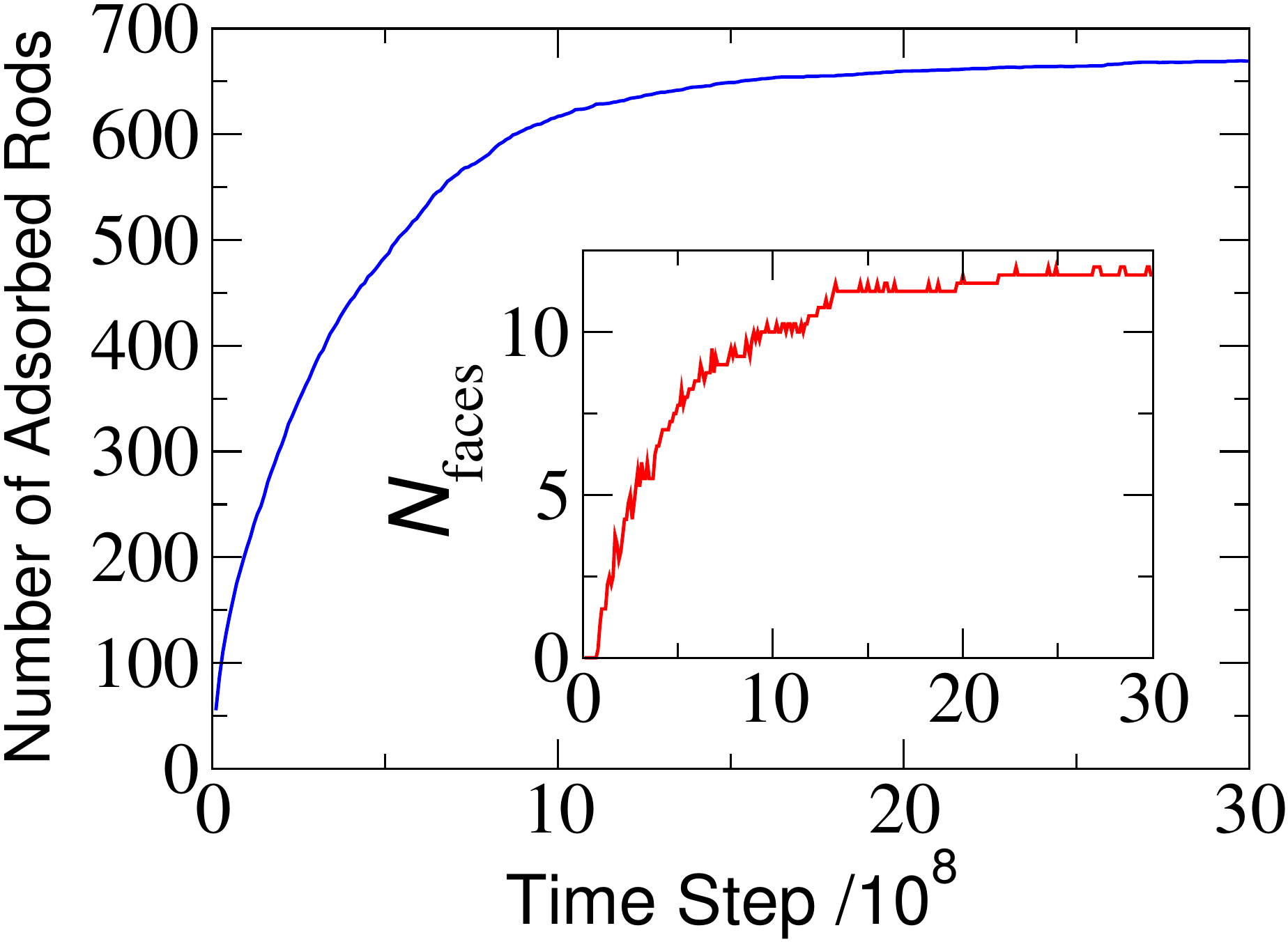}
\caption{Time evolution of the number of adsorbed rods (main) and the number of faces (inset) for a typical simulation of the spherical system, with nanoparticle radius $\Rnp=7.0$, rod length $\nb=6$, and bead-bead interaction strength $\epsilon=1$. Throughout the article, lengths are presented in units of the bead diameter $\sigmab$ and energies in units of $\kt$.}
\label{fig:equilibrium}
\end{figure}

Except where mentioned otherwise, all observables were measured after simulations had equilibrated. Equilibration was assessed by monitoring the number of adsorbed rods and the number of clusters (defined below) on the nanoparticle surface (see Fig.~\ref{fig:equilibrium} for an example). We set the total number of rods $\Ntot$ in a given system so that at most ($\sim 55\%$) of rods adsorbed onto the nanoparticle at equilibrium (ranging from $\Ntot=25$ for $\Rnp=1.0$ to $\Ntot=1080$ for $\Rnp=7.0$).  This ensured that, although the simulations were performed in the NVT ensemble, the driving force for adsorption did not diminish significantly over the course of a simulation.
The results presented were obtained by averaging over several independent simulations for each parameter value ($\sim 10$ for $ \Rnp<4.5$ and $\sim 5$ for $\Rnp>5.0$). Fewer simulations were performed for larger nanoparticles because computation times become prohibitively expensive; both the number of simulated rods and the equilibration timescale increase with nanoparticle size.

\textbf{Cylindrical model in 2D.}
To consider a case in which only mean curvature is present, we also analyze a simplified two-dimensional (2D) system, where rods are end-attached onto the outside of a circular boundary (\ie a 2D slice of a cylindrical nanoparticle). For simplicity, we do not simulate the dynamics of rod adsorption onto the cylinder. Instead, we consider a grafting density equal to close packing on the surface of the cylinder, $\Nrod=\pi/\sin^{-1}(\sigmab/2\Rnp)$, so the position of each rod end bead is fixed but rods can rotate. To investigate the large attraction strength (low temperature) limit for this system, we performed Metropolis Monte Carlo simulations \cite{Metropolis1953} in which randomly chosen rods were rotated by a small angular displacement (up to $0.01$ radians), and the well-depth was gradually increased to $\epsilon=30\kt$.  In the limit of hard spheres and a short-range attractive interaction, $\alpha \rightarrow \infty$, the energy minimizing configurations of this system can be calculated exactly. This result and typical configurations from the Monte Carlo simulations are discussed in section~\ref{sec:results}.


\section{Results}
\label{sec:results}
 Except where mentioned otherwise, we will focus on the 3D system. All of these simulations resulted in rods strongly end-adsorbed to the nanoparticle to yield a dense layer. In the absence of lateral attractions between rods ($\epsilon=0$) the bottom beads (which form the innermost layer) aquire local hexagonal order punctuated by 12 five-fold defects arranged with icosahedral symmetry, as observed for the packing of spheres on a spherical surface \cite{Bowick2000,Bausch2003}.

\begin{figure} 
\centering
\includegraphics[width=0.5\textwidth]{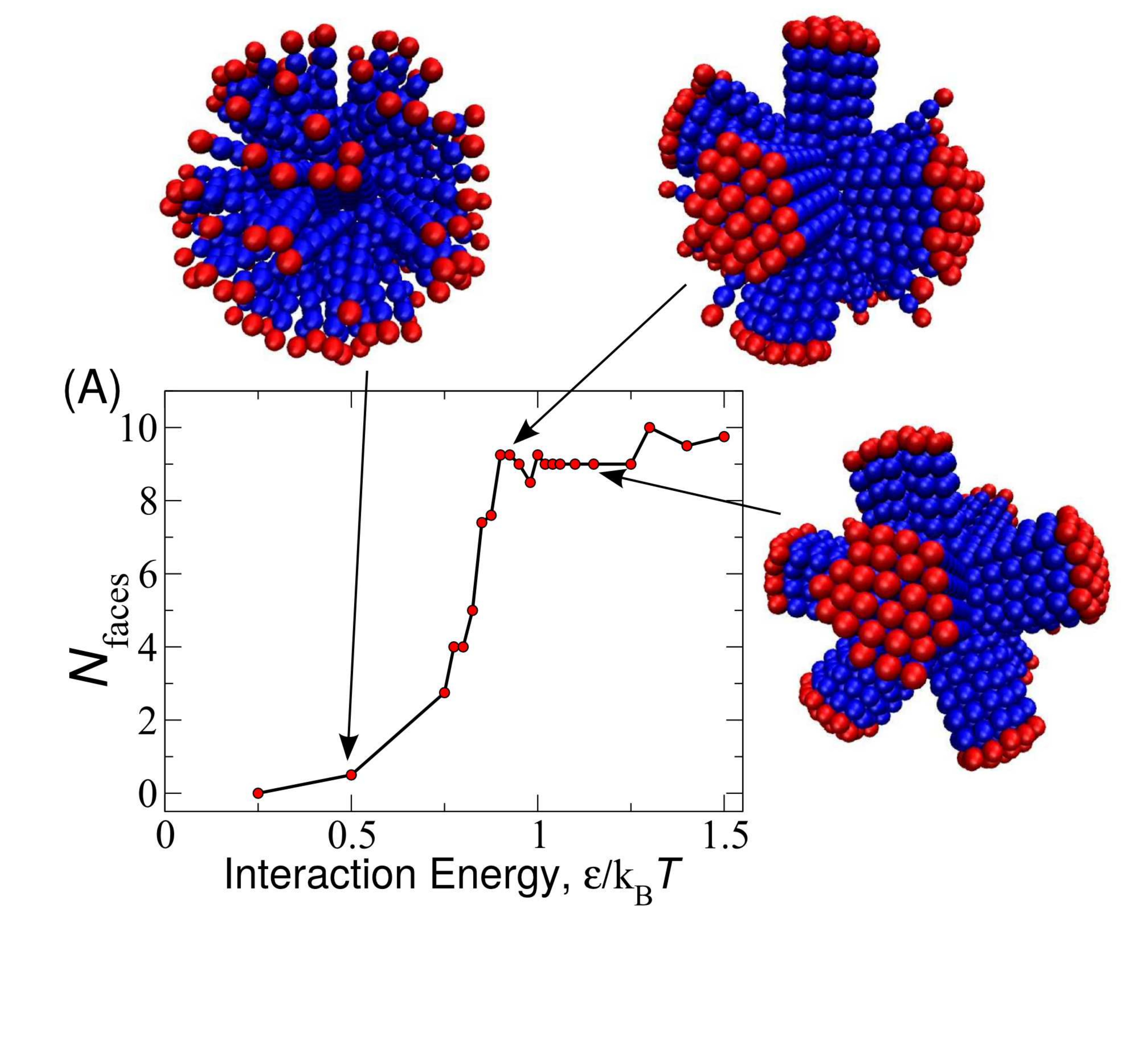}
\includegraphics[width=0.45\textwidth]{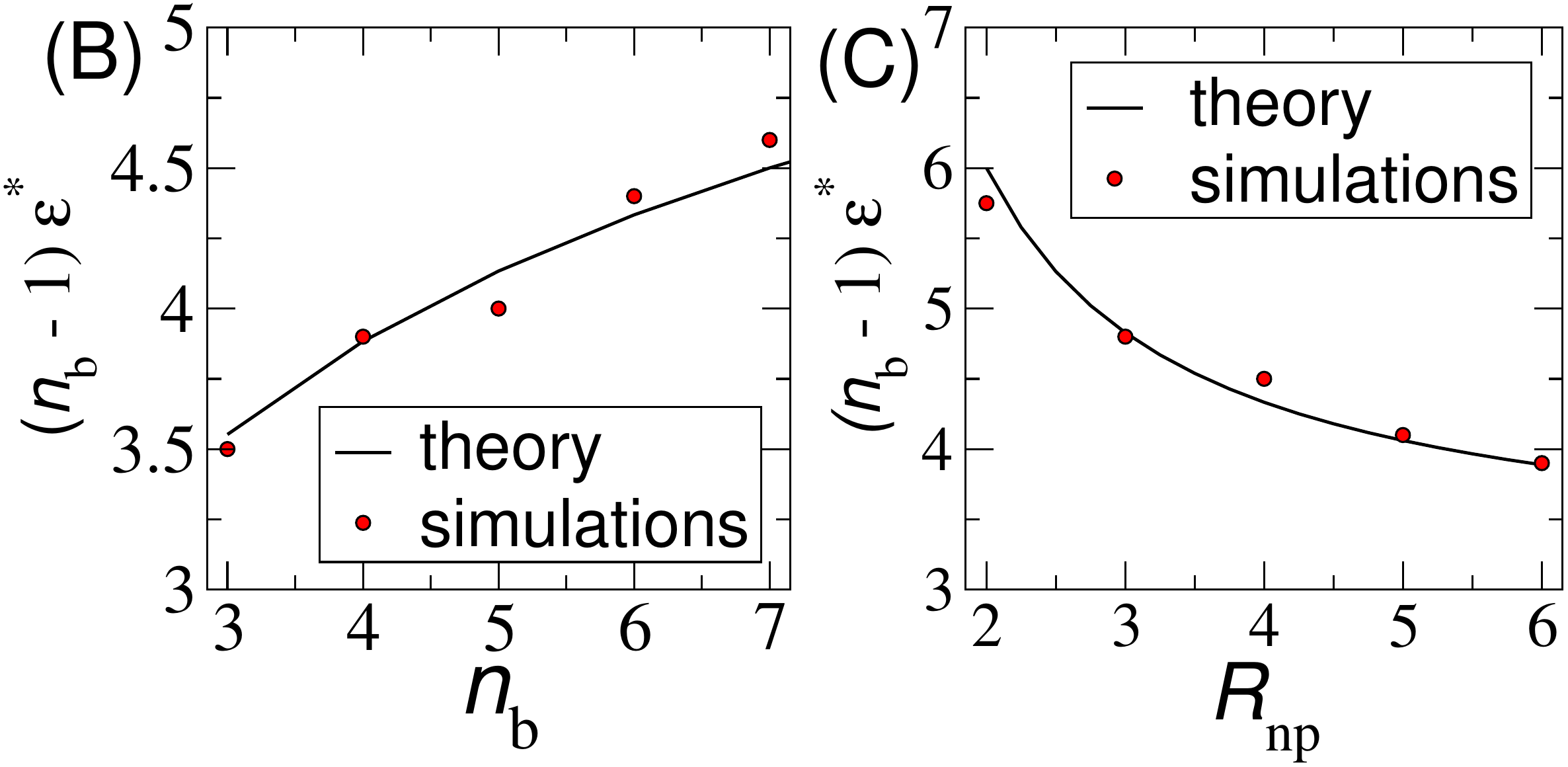}
\caption{
{\bf (A)} Number of faces $\nf$ formed as a function of the bead-bead interaction strength $\epsilon$ for rod aspect ratio $\nb{=}6$ and spherical nanoparticle radius $\Rnp{=}4$. Snapshots of final simulation structures are shown for indicated values of $\epsilon$. To aid visibility, the outermost beads are colored in red whereas others are colored blue. We determined $\epsilon^*$ as the value of the interaction strength for which  $d\nf/d\epsilon$ is maximum, calculated by fitting the results for each set of $(\nb,\Rnp)$ independently to a spline. The resulting transition interaction strengths are compared to results from the theories described in the text as a function of {\bf (B)} $\nb$ for $\Rnp=4$ and {\bf (C)} $\Rnp$ for $\nb = 6$.}
\label{fig:criticalinteraction}
\end{figure}

\textbf{A disordered-to-ordered transition occurs above a critical inter-rod interaction strength.}
For weak inter-rod interactions (small $\epsilon$), rod orientations are roughly random, leading to disordered  arrangements of beads in the layers above the surface. Above a threshold value $\epstar(\nb,\Rnp)$, we observe a transition to an ordered  layer in which rods are approximately parallel to their nearest neighbors. The value of $\epstar$ is determined by the competition between the orientational entropy of the rods in the disordered state and the attractive interactions of parallel rods (Fig.~\ref{fig:criticalinteraction}). We approximately calculate how this competition depends on $\Rnp$ and $\nb$ at the end of this section. However, we first examine the unusual configurations that arise once the system enters the ordered state.

Due to the nanoparticle curvature, rods cannot all maintain parallel orientations and fixed interaction distances with their neighbors.  Thus, the system undergoes a spontaneous symmetry breaking in which the surface is demarcated by line defects (or fissures) which form a polyhedron that inscribes the nanoparticle surface. Within each `face' of the polyhedron, rods are roughly parallel and arranged with hexagonal close packing. Interestingly, the defect lines do not necessarily form a regular polyhedron.

\begin{figure}
\centering
\includegraphics[width=0.49\textwidth]{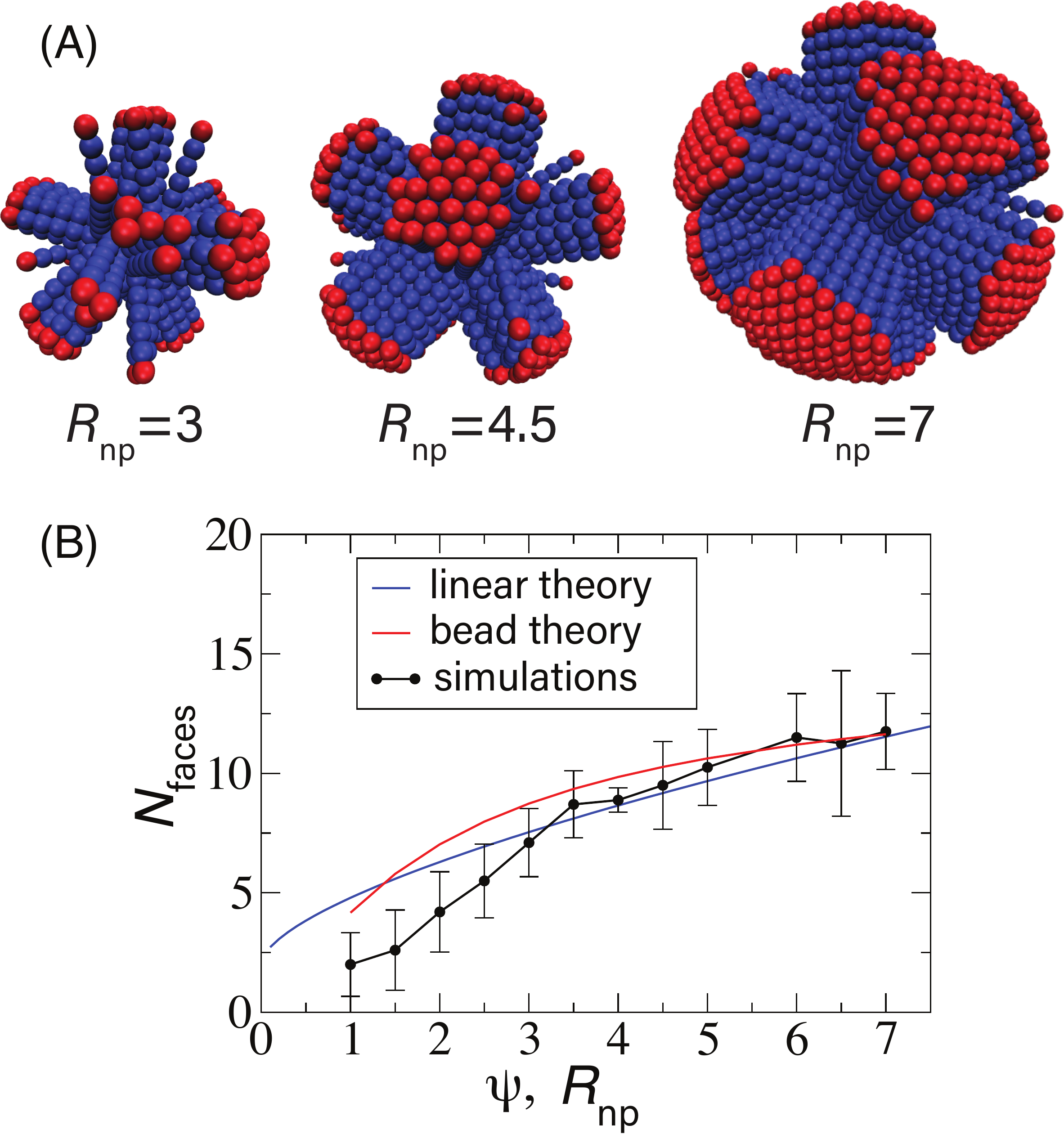}
\caption{
{\bf (A)} Snapshots of faceted structures in the spherical system shown for indicated values of the nanoparticle radius with $\nb=6$ and  $\epsilon=1$.
{\bf (B)} The number of faces is plotted as a function of the nanoparticle radius $\Rnp$ from the simulations (points) and as a function of $\psi$ for the linear tilt theory (Eqs.~\eqref{eq:linearStrain}, black line) and the theory accounting for bead intercalation (Eq.~\eqref{eq:nFaces}, red line). Error bars are estimated from 5-10 independent trials at each parameter set.
}
\label{fig:nFaces}
\end{figure}

To examine how curvature affects the surface layer geometry, we measured the number of faces as a function of nanoparticle radius. As shown in Fig.~\ref{fig:nFaces}, the number of faces increases slowly with $\Rnp$. Faces were identified by clustering neighboring rods with parallel orientations. To avoid calculating transiently disordered rods or small clusters as separate faces, we required a face to exceed a minimum cluster size $\nmin$.  The minimum cluster size must necessarily depend on nanoparticle radius, since the maximum possible size of a face decreases with $\Rnp$.  For each nanoparticle radius, we identified a local minimum in the cluster size distribution, above which there were well-formed faces and below which there were individual rods or small clusters. Based on this, $\nmin$ varied from 2 for $\Rnp = 1.0$  to 5 for $\Rnp\ge3$.

\begin{figure}[h]
\centering
\includegraphics[width=0.4\textwidth]{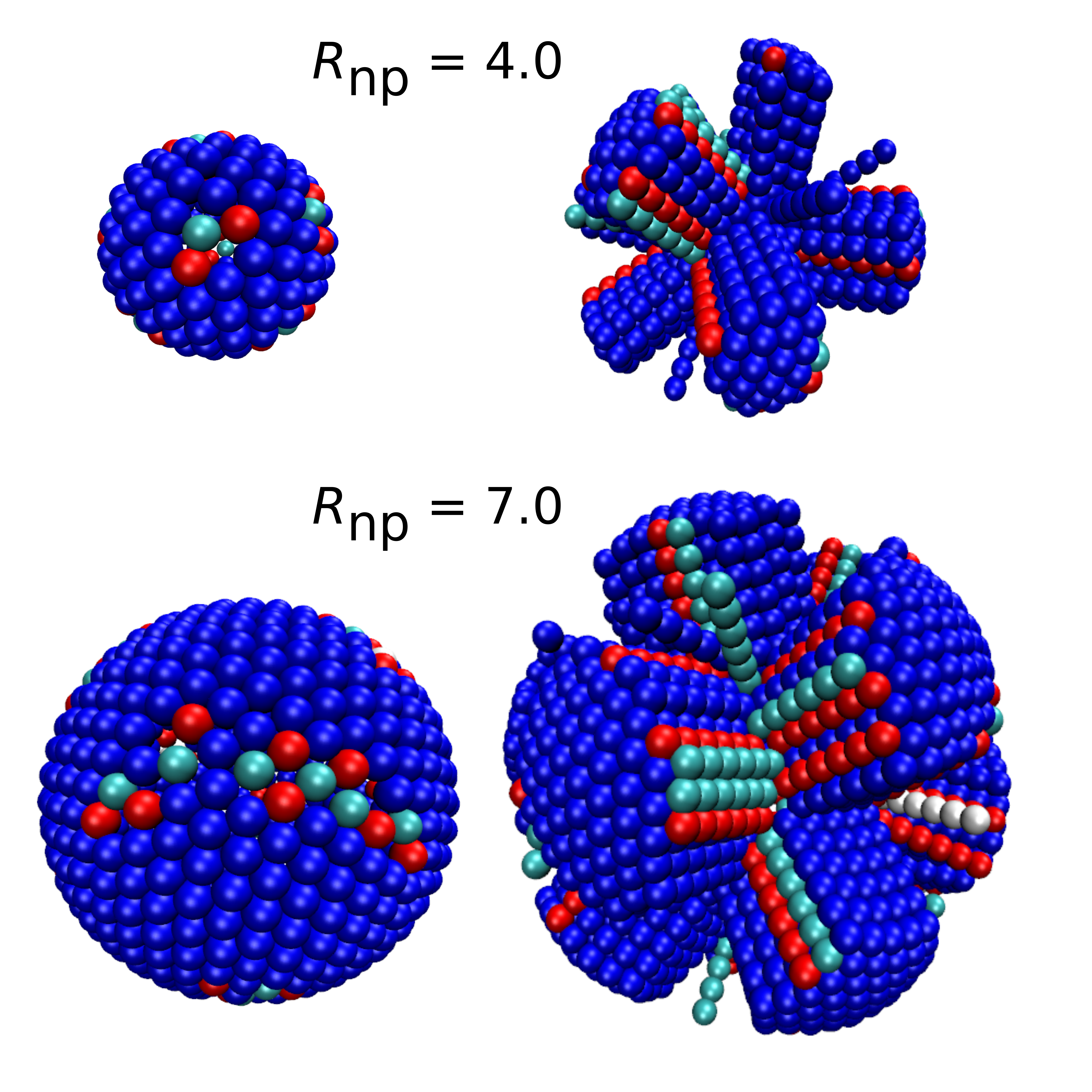}
\caption {Simulation snapshots illustrating the relationship between inner-layer disclinations and scars (left), and outer-layer fissures (right) for two different nanoparticle sizes ($\Rnp = 4.0$ (top) and $\Rnp = 7.0$ (bottom)).  The images on the left show only the inner-layer beads, while the images on the right show all beads. Each rod is colored according to the coordination number of its inner-layer bead, with colors(coordination) given by:  red(5), blue(6), cyan(7) and white(8). Coordination numbers were determined by Delaunay triangulation \cite{Renka1997,Cheong2002}.
}
\label{fig:scars}
\end{figure}

As shown in Fig.~\ref{fig:scars}, the formation of line defects in the outer layer affects the packing within the innermost layer.  Each of the 12 fivefold disclinations of the inner layer tends to be located at a vertex where several outer-layer faces meet. However, since there are typically more than 12 such vertices, not every vertex is associated with a disclination. Moreover, as the nanoparticle radius increases, we see that the disclinations extend into scars which extend along the polyhedron edges. The number and sizes of these scars are roughly consistent with the case of spherical particles adsorbing on the surface of a nanoparticle (in which case the number of excess dislocations per scar would go as $\sim 2\Rnp/\sigmab$ for $2\Rnp/\sigmab\ge5$ \cite{Bowick2000,Bausch2003}).  However, we observe that the inner-layer defects in the rod system are more disordered than those observed in the case of spherical particles.

\begin{figure}
\centering
\includegraphics[width=0.49\textwidth]{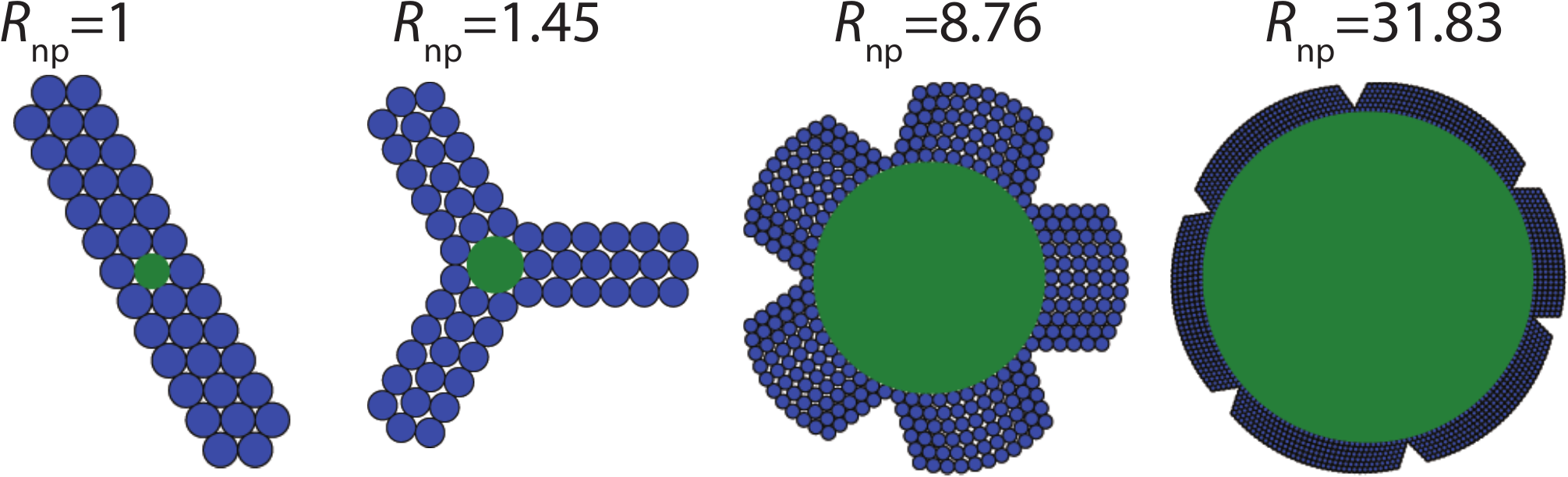}
\includegraphics[width=0.35\textwidth]{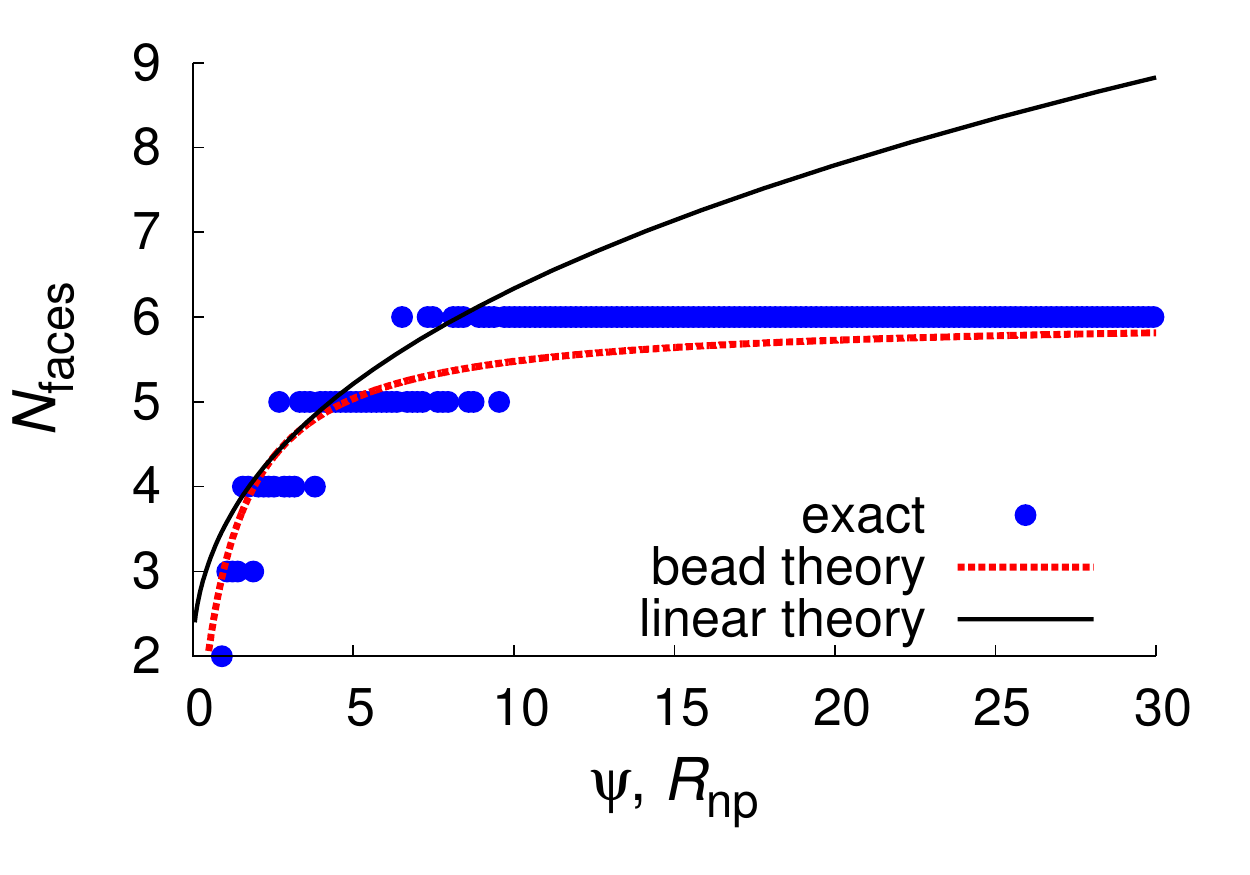}
\caption{
The number of faces in a circular geometry is plotted as a function of $\Rnp$ for results from exact energy minimizing configurations (points) and the bead intercalation theory (red dotted line), and as a function of  $\psi$ for the theory (Eqs.~\eqref{eq:linearStrainCyl}, line). Snapshots from Monte Carlo simulations at several values of $\Rnp$ are shown on top, with rods colored blue and the nanoparticle in green.
}
\label{fig:nFacesCyl}
\end{figure}

Despite the evident coupling between the locations of the inner-layer disclinations and the fissures in the outer layer, we find that the Gaussian curvature responsible for the former is not required for the formation of the latter.  In particular, simulations on the 2D cylinder system (which has only mean curvature) exhibited analogous fissures in the outer layer.  Example low energy configurations for several nanoparticle radii are shown in Fig.~\ref{fig:nFacesCyl}.


\textbf{Factors that determine the number of faces.}
It is well known that incompatibility between a material's preferred local packing geometry and long-range order can lead to defect-stabilized phases (\eg Frank-Kasper phases, blue phases, and twist-grain-boundary phases), in which finite-sized unfrustrated domains are separated by regions of broken short-range order (defects) \cite{Kleman1989}. 
 In our simulations, we observe an analog of this behavior on finite nanoparticle surfaces. The resulting defect geometries can be understood from the following simple models.\\

\begin{figure}[h]
\centering
\includegraphics[width=0.225\textwidth]{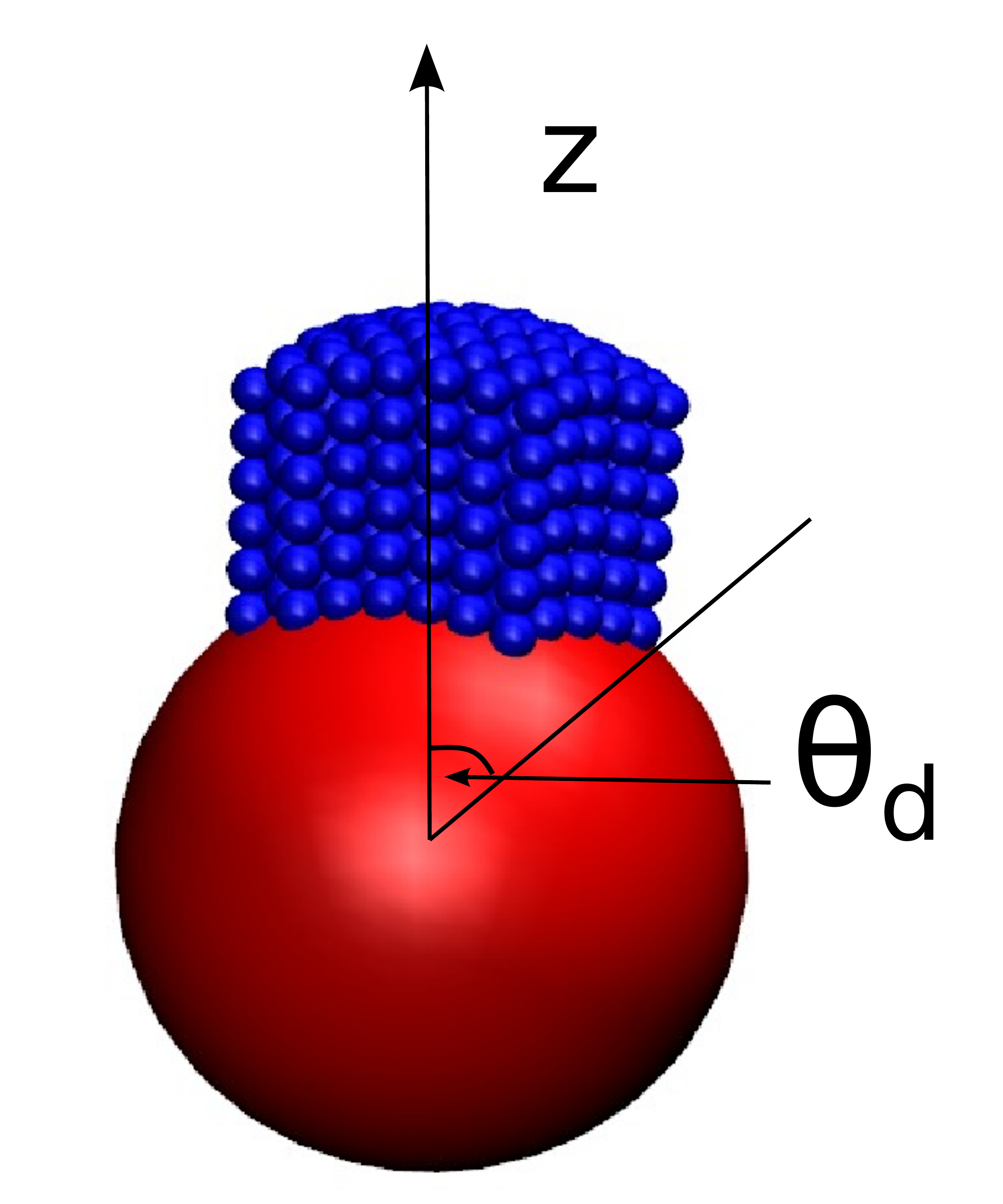}
\caption{Schematic of a ``face'', defined as a cluster in which the constituent rods are nearly parallel. The base of the cluster in contact with the nanoparticle defines a patch that is bounded by an angle $\thetac$.}
\label{fig:schematicangle}
\end{figure}

\textit{ Bead Theory:}
We consider a face (patch of nearly parallel, hexagonally ordered rods) whose orientations are aligned with the sphere normal at the patch center, which we define as the z-axis (Fig.~\ref{fig:schematicangle}). Away from the center of the face, the nanoparticle curvature forces rods out of register.  For the rod excluded-volume geometry that we consider (a chain of spheres), this displacement can be accommodated by intercalation of neighboring rods until the patch extends along the surface to an angle bounded by
\begin{align}
\thetac \approx \pi/6+0.5\sigmab/\Rnp
\label{eq:beadTheory},
\end{align}
where the second factor approximately accounts for rod granularity.
Growth of the patch beyond $\thetac$ would lead to a large strain energy which increases with patch size. This is avoided by terminating the  patch and commencing a new patch with a different rod orientation. The two patches are thus separated by a line defect.
The number of faces $\nf$ can then be estimated from the fraction of the nanoparticle surface occupied by a single patch, giving
\begin{align}
\nf = &2/(1-\cos\thetac)  \quad \mbox{for spherical nanoparticles} \label{eq:nFaces} \\
\nf = &\pi/\thetac \qquad \mbox{for cylindrical nanoparticles}
\label{eq:nFacesCyl}.
\end{align}
These relationships qualitatively agree with the numbers of faces observed in simulations (Figs.~\ref{fig:nFaces} and ~\ref{fig:nFacesCyl}).

For the 2D cylinder system, the bead theory exactly calculates the energy minimizing configurations in the limit of hard spheres and a short-range attractive interaction, $\alpha \rightarrow \infty$.   The number of faces is given by
\begin{align}
\nf^\text{exact}=\text{ceiling}\left(\Nrod/\text{floor}(2+\Nrod/6)\right)
\label{eq:nfexact}
\end{align}
where the `floor' and `ceiling' operands account for the fact that the number of domains and the number of rods within a domain must be integers. Due to these constraints $\nf^\text{exact}$ switches between neighboring values as the nanoparticle radius varies (Fig.~\ref{fig:nFacesCyl}) and the resulting dependence of $\nf^\text{exact}$ on nanoparticle radius is shown in Fig.~\ref{fig:nFacesCyl}. We confirmed the prediction of Eq.~\eqref{eq:nfexact} by comparison with Monte Carlo simulations (see section~\ref{sec:model}).

\vspace{1 cm}
\textit{Linear Theory:}
 While the model just described depends on the local packing geometry of our rods, the effect is general. To show this, we consider deposition on a spherical or cylindrical substrate of a film with sufficient thickness $\delta$ such that shear or tilt deformations correspond to lower energy than bending. For simplicity, we consider a linear tilt modulus $G$ (although our simulated rod monolayer has a highly nonlinear stress-strain relationship).

 We then consider a circular domain of such a material as defined above. As the domain extends across the surface by an angle $\thetac$, the material undergoes a strain along $\hat{\mathbf{z}}$ given by $\gamma(\theta)=\tan\theta$. The domain boundary gives rise to an interfacial energy characterized by a line tension $\sigma$, which we expect to be proportional to the film thickness $\delta$. The total energy density $\Utot$ is given by integrating the strain energy over the domain area, and summing over the number of domains $\nf$:
\begin{align}
\frac{\Usphere(\thetac)}{4 \pi \Rnp^2 \delta G}=&\frac{\nf(\thetac)}{2}\left[ \left(\cos\thetac + \sec\thetac-2\right) + \psi^{-1} \sin \thetac\right]
\label{eq:linearStrain}
\end{align}
with
$\psi\equiv \Rnp \delta G/\sigma$ the dimensionless ratio between shear modulus and line tension. The term in parentheses on the right-hand side of Eq.\eqref{eq:linearStrain} gives the strain energy, the next term is the interfacial energy, and we neglect contributions from the 12 disclinations required by the spherical geometry 
\footnote{One might also consider rod-like particles interacting through depletion interactions, such as in the colloidal membranes system studied by Barry et al. \cite{Barry2010}. Since the depletion interaction depends only on overlapping occluded volume rather than pairwise interactions between specific molecular groups, tilting affects the interaction between neighboring rods only near their ends. However, since neighboring rods cannot overlap, tilting reduces the aerial density by a factor $\rho(\theta) \propto 1-\cos(\theta)$. Accounting for this density reduction leads to a similar dependence of $\nf$ on $\Rnp$ as in Eq.~\eqref{eq:linearStrain}.}.
Because the tilt modulus and interfacial energy are both dictated by the strength of molecular interactions, we
expect the line tension to be proportional to the film thickness $\delta$; thus, $\psi \propto \Rnp/\sigmab$ with $\sigmab$ the  molecular spacing.
A similar analysis for a cylinder with radius $\Rnp$ and length $L$ yields
\begin{align}
\frac{\Ucyl(\thetac)}{2 \pi \Rnp L \delta G}=&\frac{1}{2\thetac}\left[ \left(\tan\thetac-\thetac\right) + \psi^{-1}\right]
\label{eq:linearStrainCyl}
\end{align}
The size and the number of domains $\thetac$ and $\nf$ are obtained by minimizing Eq.\eqref{eq:linearStrain} or \eqref{eq:linearStrainCyl}. As shown in Fig.\ref{fig:nFaces}, the linear theories qualitatively describe the simulation results, except that the predicted number of faces monotonically increases with $\Rnp$. In the simulated systems, the bead intercalation cuts off $\nf$ at about $15$ or $6$ for spheres or cylinders respectively, and thus the bead theory is more quantitatively accurate for large $\Rnp$. It is difficult to see this cut off in the sphere geometry because simulations became computationally intractable for large $\Rnp$, but the effect is clear in the cylindrical geometry. In both geometries the number of domains approaches $2$ as $\psi$ approaches zero (corresponding to small particle size or small tilt modulus). In this limit, the material in each hemisphere aligns with the z-axis with an equatorial defect between the two domains, resembling structures observed in simulation of alkanes on gold nanoparticles \cite{Lane2010, Luedtke1998}. The similarity between the results on spherical and cylindrical substrates emphasizes that the frustration associated with depositing a thick film on a curved substrate is correlated to mean curvature, in contrast to the frustration of packing discs on a spherical surface associated with Guassian curvature.

{\bf An upper critical nanoparticle size for faceting.} For sufficiently large nanoparticles the system will enter the thin film limit, with bending deformations lower in energy than tilt. (While $\Usphere$ and $\Ucyl$ monotonically increase with $\Rnp$, the bending energy of an elastic layer is independent of $\Rnp$ in a spherical geometry and decreases as $\Rnp^{-1}$ for a cylinder.) We thus anticipate a critical nanoparticle size above which faceting does not occur, given by $\Rnp^{*} \sim \delta \sigmab \alpha$ with $\alpha^{-1}$ the width of the interparticle potential.

{\bf Intra-bundle twist leads to another mechanism of frustration:} While the interactions along the rods are uniform in the simulations described to this point (Model I, Fig.~\ref{fig:schematic}), simulations with interactions confined to the middle two beads (Model II) resulted in an approximately regular dodecahedron. The $12$ faced morphology arises because the rods form twisted bundles, which presumably are  allowed due to the increased orientational freedom available to rod lateral interactions when the attractions are limited to the two middle beads. Inter-filament twist provides another form of frustration which can limit the lateral growth of bundles \cite{Grason2007, Yang2010, Grason2008,Grason2009}, which likely couples with the curvature induced frustration to decrease the critical face size.

\begin{figure}[h]
\centering
\includegraphics[width=0.4\textwidth]{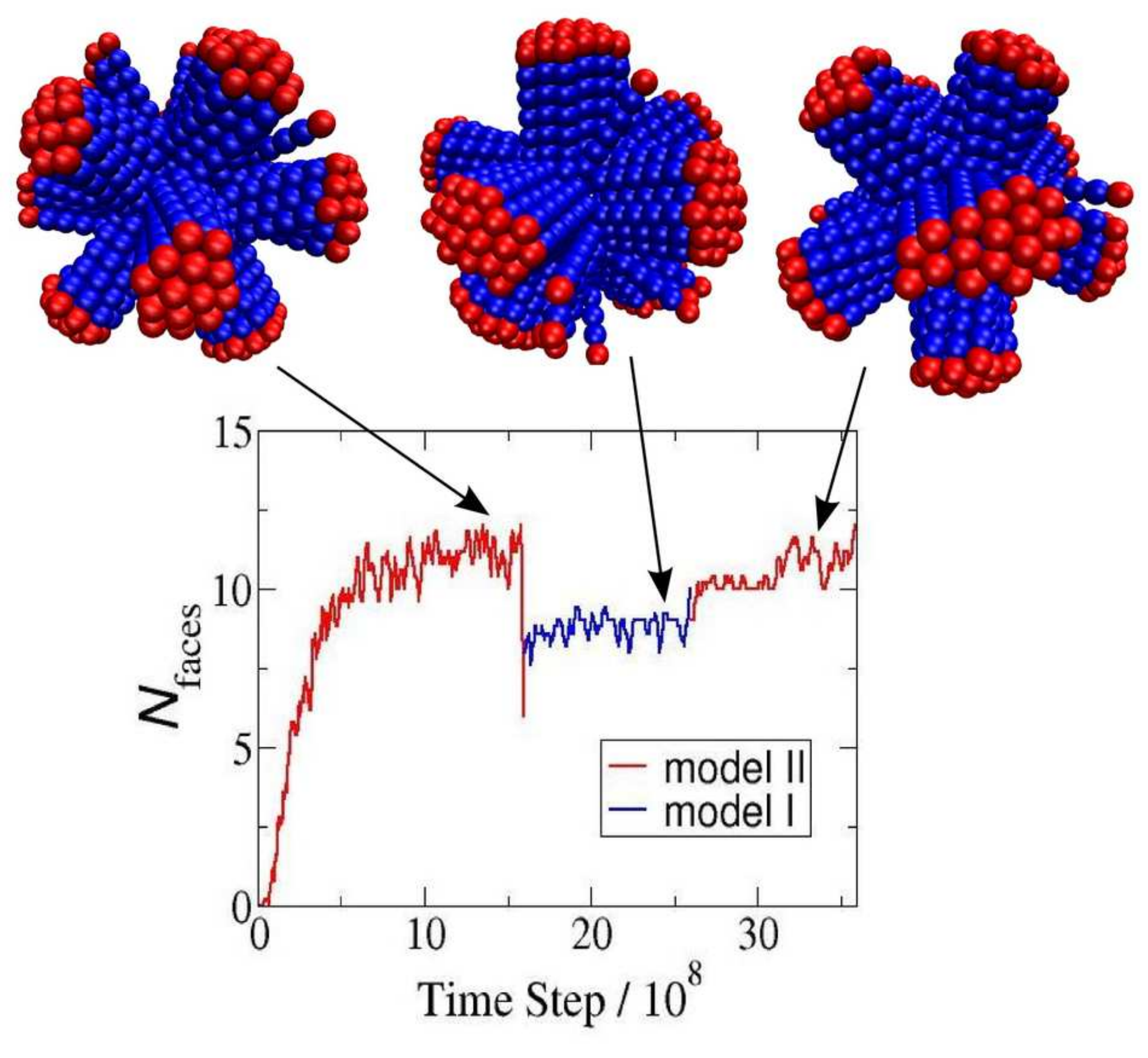}
\caption{
Reversible transitions between morphologies: the number of faces is shown as a function of simulation time step, for a simulation starting with the Model II potential, switching to the Model I potential, and returning to the Model II potential.  The time series demonstrates that the simulation outcomes are robust to changes in initial condition.
}
\label{fig:transition}
\end{figure}

Given the magnitude of structural difference which arises between a seemingly small change in the interaction potential, we decided to further test whether these are equilibrium configurations.  We performed an additional set of simulations in which the system was first equilibrated under Model II interactions, then switched to Model I interactions and re-equilibrated, and then finally switched back to Model II interactions.  Note that this simulation protocol is not meant to model a particular experiment (in which it might be difficult to switch interactions \emph{in situ}), but rather to test whether the simulation outcomes depend on initial condition.   As shown in Fig.~\ref{fig:transition}, the $12$-face structure obtained from Model II interactions reconfigures to a $9$-face structure upon switching to Model I interactions, and then returns to the $12$-face structure upon switching back to Model II.  Since these reconfigurations require extensive morphological changes, these reversible changes suggest that the structures indeed correspond to equilibrium morphologies.

\textbf{Calculation of disorder-to-order transition energy.}
Finally, we consider how the order/disorder transition interaction strength $\epstar$ depends on the size of the nanoparticle and the rod aspect ratio.
The formation of order is determined by the competition between the orientational entropy of rods in the disordered state and the attractive interactions of rods in the polyhedral state. The orientational free energy of a rod in the disordered state is given by $\ffree/\kt = \log 2$, where we neglect excluded volume with neighboring rods, assume that the nanoparticle surface restricts rod orientations to a solid angle of $2\pi$, and that rods have free orientations in the reference state. To calculate the free energy of rods within a polyhedral face, we first estimate the rod-rod binding free energy from the partition function for the bound state of two neighboring rods:
\begin{align}
\beta \fbind(\epsilon,\nb)= & -\log [q_\text{rod}(\epsilon,\nb)/4 \pi] \nonumber \\
q_\text{rod}(\epsilon,\nb) = &\frac{2 \pi}{\sigmab^2 \beta V_\text{morse}^{''} \sum_{n=1}^{\nb-1} n^2 }  \exp[(\nb-1) \epsilon/ \kt]
\label{eq:fbind}
\end{align}
with $q_\text{rod}$ the partition function for the interaction between two neighboring rods integrated over angular fluctuations, $V_\text{morse}^{''}=2 \epsilon \alpha^2$ as the second derivative of the Morse potential evaluated at its minimum, and we have expanded the interaction potential to second-order. Notice that the relevant rod-rod interaction strength parameter is $(\nb-1)\epsilon$, since the interaction between the innermost beads of neighboring rods is independent of their orientations.
Each interior rod within a polyhedral face interacts with six neighbors, whereas those on the perimeter have on average $3.5$ neighbors.  We therefore approximate the binding free energy of interior rods as $\fbind(3 \epsilon)$, accounting for double counting, and perimeter rods as $\fbind(3.5 \epsilon/2)$.  For a face with $\Nrod$ rods, the fraction of rods on the perimeter can be approximated as $\fperim = 2 (\pi/\Nrod \rho_\text{HCP})^{1/2}$, with $\Nrod = 4 \pi \Rnp^2 \rho_\text{HCP}/\sigmab^2\nf(\Rnp)$,  $\rho_\text{HCP}=2/\sqrt{3}$ and $\nf(\Rnp)$ the number of faces as a function of
nanoparticle size, given by Eqs.~\eqref{eq:beadTheory} and \eqref{eq:nFaces}.  The free energy per rod within a face is then given by
\begin{align}
\frod(\epsilon,\nb, \Rnp)/\kt  & = \left(1-\fperim\right) \fbind(3 \epsilon, \nb) \nonumber \\
& + \fperim \fbind(1.75 \epsilon, \nb)
\label{eq:fface}
\end{align}
Finally, the value of the transition $\epstar$ is determined by $\frod(\epstar, \nb, \Rnp) = \ffree$.  The calculated transition values are compared to simulation results in Fig.~\ref{fig:criticalinteraction} (B, C) as functions of $\nb$ and $\Rnp$.  We see that the theoretical prediction agrees closely with the simulation results except at small $\Rnp$, where the continuum limit breaks down.

\section{Conclusions}
In conclusion, our simulations and simple theoretical models demonstrate that the geometric frustration intrinsic to
assembling anisotropic particles on curved surfaces leads to diverse equilibrium morphologies. While the specific
number of faces obtained at a given parameter set depends on the details of the interparticle interaction potential, a
simple model based on a generic linear stress-strain relationship predicts qualitatively similar behaviors.  Thus, it
should be possible to generate polyhedral morphologies using a wide range of particle shapes and interactions.
Extending our model to describe other interparticle interactions or substrate geometries would enable designing
alternative assembly morphologies. Controlling the strength of interparticle interactions (e.g. by temperature
or depletant concentration) enables reversible switching between ordered, faceted morphologies and disordered structures.
Moreover, by changing the form of inter-particle potentials, it is possible to obtain highly symmetric structures (\eg a dodecahedron) or
asymmetric structures (\eg the 9-face polyhedron).

Our model was particularly motivated by two experimental systems. First, Dogic and coworkers \cite{Barry2010} showed that a suspension of rodlike viruses and non-adsorbing polymer form 2D colloidal membranes, comprised of one rod-length thick monolayers of aligned rods. Deposition of such a monolayer onto a colloidal particle could be driven by depletion interactions \cite{Asakura1954} or complementary functionalization of rod ends and nanoparticle surfaces. It would be interesting to observe the morphologies of such assemblages as a function of nanoparticle size, depletant concentration, and depletant size. Second, cryoEM micrographs of HIV GAG proteins assembled around functionalized nanoparticles exhibit extensive scars that resemble features of the line defects and disorder observed in our simulations (Fig.~\ref{fig:scars}).
 However, we note that the GAG protein assembles with a preferred curvature in the absence of a spherical template, in contrast to the rod-like particles considered here. The physics described here will be most relevant to cases in which the nanoparticle radius is small in comparison to the cone angle. For cone angles which are commensurate with the nanoparticle radius, the frustration resulting from mean curvature would be avoided, and we would expect similar morphologies to those observed for packing of discs on a sphere \cite{Chen2007a}.

\begin{acknowledgments}
We thank Greg Grason for insightful comments on the manuscript. This work was supported by Award Number R01GM108021 from the National Institute Of General Medical Sciences and the Brandeis Center for Bioinspired Soft Materials, an NSF MRSEC, DMR-1420382. Computational resources were provided by the NSF through XSEDE computing resources (Trestles and Stampede) and the Brandeis HPCC which is partially supported by the Brandeis MRSEC, DMR-1420382. MFH wrote part of this manuscript while at the Aspen Center for Physics, which is supported by NSF grant PHY-1066293.
\end{acknowledgments}



%

\end{document}